\documentclass[12pt,preprint]{aastex}

\usepackage{graphicx,natbib,color}

\shorttitle{Onsets and spectra of impulsive solar energetic particles}
\shortauthors{E.P. Kontar and H. A. S. Reid}

\begin{document}


\title{Onsets and spectra of impulsive solar energetic electron events observed near the Earth}


\author{Eduard P. Kontar and Hamish A. S. Reid}
\affil{ Department of Physics and Astronomy,
University of Glasgow, G12 8QQ, United Kingdom}
\email{eduard@astro.gla.ac.uk, hamish@astro.gla.ac.uk}




\begin{abstract}
Impulsive solar energetic electrons are often observed in the interplanetary space
near the Earth and have an attractive diagnostic potential for poorly understood
solar flare acceleration processes. We investigate the transport of solar flare
energetic electrons in the heliospheric plasma to understand
the role of transport to the observed onset and spectral properties
of the impulsive solar electron events. The propagation of energetic electrons
in solar wind plasma is simulated from the acceleration region at the Sun
to the Earth, taking into account self-consistent generation and absorption
of electrostatic electron plasma (Langmuir) waves,
effects of non-uniform plasma, collisions and Landau damping.
The simulations suggest that the beam-driven plasma turbulence
and the effects of solar wind density inhomogeneity play
a crucial role and lead to the appearance of a) spectral break
for a single power-law injected electron spectrum,
with the spectrum flatter below the break, b) apparent early onset of low-energy
electron injection, c) the apparent late maximum
of low-energy electron injection. We show that the observed onsets,
spectral flattening at low energies, and formation of a break energy at tens of keV
is the direct manifestation of wave-particle interactions in non-uniform
plasma of a single accelerated electron population with an initial power-law spectrum.
\end{abstract}



\section{Introduction}

Solar flares are extremely efficient at accelerating electrons to non-thermal energies,
which can subsequently be observed either by their emission at X-ray
and radio wavelengths or escaping along open magnetic field lines
via direct electron measurements near the Earth \citep[see][for a review]{Aschwanden02,BrownKontar05,Benz08}.
The first in-situ observations of energetic particles \citep{vanAllen65}
opened up the non-electromagnetic window of flare accelerated particle observations.
Solar impulsive electron events detected in-situ generally display broken
power-law energy distributions with lower energies having softer spectra \citep{Lin85}.
These events also show near time-of-flight velocity dispersion and a beamed pitch-angle distribution \citep[e.g.][]{Lin85,Krucker_etal1999,Krucker_etal07}.
From this evidence, it is often believed that such electrons propagate
scatter-free from the Sun to the Earth \citep[e.g.][]{Wang_etal06}.
The observed correlation between the spectral indices of energetic electrons
at the Sun from X-ray data and the Earth from in-situ data
\citep{Lin85,Krucker_etal07} is often viewed as an additional support
for this model. At the same time, impulsive solar energetic electron events
are closely related observationally
\citep[e.g.][]{Li_etal1981,Ergun_etal98,Gosling_etal2003,Cane2003,Krucker_etal07}
and theoretically \citep{GZ1958,Zaitsev_etal72,Grognard1982,Melrose90,Melnik99,Kontar01a,Ledenev04}
to Type III solar radio bursts.

The standard model of Type III solar radio bursts \citep{GZ1958}
suggests that electron beams propagating in the ambient
solar wind plasma from the Sun to the Earth can excite Langmuir waves,
which in turn generate escaping plasma radio emission \citep[see][for a review]{Melrose_85}.
In a one-dimensional treatment assuming travel along magnetic field lines,
as faster electrons outpace slower electrons a positive gradient is formed in velocity space.
If this positive gradient gets large enough to start the generation of waves,
electron energy is resonantly transferred to Langmuir waves in the background plasma.
This transfer of energy reduces the gradient in velocity space forming a plateau \citep{Vedenov_etal1962,Drummond_Pines1962}. For a spatially limited electron beam cloud,
the plasma waves generated at the front of the cloud are absorbed
by the electron beam at the back allowing the electrons to travel through the corona
with small energy losses and with a velocity that decreases with time
due to plasma inhomogeneity \citep{Kontar01a}. Although this broad picture
is often supported by observations, the detailed picture of electron transport
and plasma radio emission is far from well-understood. This is largely due to electron
beam propagation and radio emission
being essentially a non-linear multi-scale problem, and is the subject
of a large number of ongoing simulation efforts
\citep[e.g.][]{Melnik99,Kontar01a,KontarPecseli02,Ledenev04,Li_etal2006,2008ApJ...677..676G}.

Solar impulsive electron events can span a broad range of energies,
from a few keV to hundreds of keV \citep{Lin_etal1996}. Their energy distribution
forms a broken power-law spectrum with the break energy in deca-keV range.
Despite often showing the near time-of-flight dispersion, lower energy electrons
appear to arrive sooner than expected from a scatter-free model \citep{Wang_etal06}.
Since the low energy electrons of a few keV should lose their energy collisionally
in the low corona, these electrons are believed to be accelerated high in the corona \citep{Lin_etal1996}.
Recent time-of-injection analysis \citep{Wang_etal06} assuming scatter-free propagation
of solar energetic electrons suggests the existence of two electron populations,
one low energy beam injected before the start of the type III burst and
one high energy beam injected after the type III burst.

In this Letter, we investigate the electron propagation from the Sun to
the Earth taking into account the scattering of electrons by beam-driven
plasma waves. We show, for the first time, that the generation and absorption
of Langmuir waves by an electron beam in the non-uniform solar corona
leads to the appearance of a break energy in the observed spectrum
at the Earth and naturally explains the observed apparent early injection
of low energy electrons.

\section{Modelling electron propagation in the heliosphere}
The transport of energetic electrons in the heliospheric plasma
is governed by a variety of different processes \citep[see][for a review]{Melrose90}.
In this work we consider solar energetic electrons propagating
along open magnetic field lines and assume their transport
can be described one-dimensionally ignoring electromagnetic effects.
Under this assumption, the evolution of the electron distribution
function $f(v,x,t)$ [electrons cm$^{-4}$ s] and the spectral energy density of electron plasma waves
$W(v,x,t)$ [ergs cm$^{-2}$] can be described self-consistently by the following kinetic equations
\cite[e.g.][]{Kontar01a}
\begin{equation}
\frac{\partial f}{\partial t} + v\frac{\partial f}{\partial x} = \frac{4\pi ^2e^2}{m^2}\frac{\partial }{\partial v}\frac{W}{v}\frac{\partial f}{\partial v}
\label{eqk1}
\end{equation}
\begin{equation}
\frac{\partial W}{\partial t} + \frac{\partial \omega}{\partial k}\frac{\partial W}{\partial x} -\frac{\partial \omega _{pe}}{\partial x}\frac{\partial W}{\partial k}  =  (\gamma(v,x) - \gamma_c - \gamma_L )W
\label{eqk2}
\end{equation}
where $\partial \omega/\partial k = 3v_{Te}^2/v$ is the group velocity of Langmuir waves, $k$ is the wavenumber
of a Langmuir wave, $\gamma (v,x) =\frac{\pi \omega_{pe}}{n}v^2\frac{\partial f}{\partial v}$
is the plasma wave growth rate, and $\gamma_c$ and $\gamma_L$ are the collisional and
Landau damping rates of waves respectively.
The first term on the right hand side of both Equation (\ref{eqk1}) and (\ref{eqk2}) describes
the resonant interaction, $\omega_{pe}=kv$, of electrons and Langmuir waves and
was first derived by \citet{Vedenov_etal1962} and \citet{Drummond_Pines1962}.
$W(v,x,t)$ is normalized to the wave energy density $\int W dk$ [ergs cm$^{-3}$] and plays a similar
role for plasma waves as the electron distribution function does
for particles.

\subsection{Initial electron beam distribution}

The initial distribution function is assumed to be a power-law
with spectral index $\alpha$ in velocity space and has a finite
spatial size $d$ at initial time $t=0$:
\begin{equation}
f(v,x,t=0) = g_{o}(v)\exp\left(-\frac{x^{2}}{d^{2}} \right)
\label{init_f}
\end{equation}
where
\begin{equation}
g_0(v) = n_{beam}\frac{(\alpha -1) }{v_{min}}\left(\frac{v_{min}}{v}\right)^{\alpha},\;\;\; \alpha > 1
\label{init_g0}
\end{equation}
is the initial electron distribution function normalized to $n_{beam}$, the beam electron
number density, $v_{min}$ is the low velocity cutoff, and $\alpha$ is the spectral
index of the initial electron beam. The injected electron flux density differential
in energy $F_0(E,x,t=0)$ [electrons cm$^{-2}$ keV$^{-1}$ s$^{-1}$] is
also a power law $F_0(E)\sim E^{-\delta}$, where $\delta=\alpha/2$.
The initial spectral energy density of the Langmuir waves is assumed
to be thermal $W(v,x,t=0) = k_BT/(2\pi^2\lambda_D^2)$,
where $T$ is the background plasma temperature, $k_B$ is Boltzmann constant
and $\lambda_{De}$ is the electron Debye length.

\subsection{Heliospheric plasma density}

The background plasma is modelled using a heliospheric density model
by \citet{parker58} with normalization by \citet{Mann_etal1999}
that can be found by numerically integrating
the equations for a stationary spherical symmetric solution
for solar wind \citep{parker58}
\begin{equation}\label{sol1}
r^2n(r)v(r)= C= const
\end{equation}
\begin{equation}\label{sol2}
  \frac{v(r)^2}{v_c^2}-\mbox{ln}\left(\frac{v(r)^2}{v_c^2}\right)=
  4\mbox{ln}\left(\frac{r}{r_c}\right)+4\frac{r_c}{r}-3
\end{equation}
where $v_c\equiv v(r_c)=(k_BT/\tilde{\mu}m_p)^{1/2}$,
$r_c=GM_s/2v_c^2$, $T$ is the electron plasma temperature, $M_s$ is the
mass of the Sun. The constant appearing above is fixed by satellite measurements
near the Earth's orbit (at $r = 1$~AU, $n =6.59$~cm$^{-3}$) and equates
to $6.3\times 10^{34}$~s$^{-1}$.

\section{Simulation of electron transport}

The system of kinetic equations (\ref{eqk1}, \ref{eqk2}) have been solved using
finite difference methods as described in \citep{Kontar01a} for a variety
of initial beam parameters.  The temperature of the heliospheric plasma was taken to be
constant at $T=10^6K$ and the plasma density profile is given by the numerical
solution of Equations (\ref{sol1}, \ref{sol2}). The low velocity cutoff was taken equal to approximately twice
the thermal electron velocity $v_{min}=1.2\times 10^9$~cm/s.
The initial spatial size of the electron cloud was taken as $d=5\times 10^9$~cm
so the injection time of electrons with velocity $5\times 10^9$~cm/s is one second,
which is a typical duration of type III bursts near the starting frequencies
\citep{Dulk85}.  The initial beam density $n_{beam}$ was varied
from $1\times 10^{-3}$~cm$^{-3}$ to $1$~cm$^{-3}$ to explain
the observed fluxes near the Earth. The fluence spectra of such beam densities 
correspond to weak to medium solar energetic electron events observed near the Earth 
and are required for realistic simulation computational times.
The initial height of the beam was $5\times 10^9$~cm corresponding to the density
$2.1\times 10^9$~cm$^{-3}$ (local plasma frequency $\sim 415$~MHz).
The initial beam spectral indices $\delta=\alpha/2$ were between
$2.5$ and $4.5$, consistent with the
observational values \citet{Krucker_etal09}.

\subsection{Electron spectra at 1AU}
Generally, it is seen from our simulations that as soon as the plasma wave growth
time $\sim 1/\gamma (v,x)$ is less than the time scale of an electron cloud $d/v$
for some energy $E(v)$, the wave-particle interactions start to play an important
role in the electron transport. Since this condition is energy dependent
the electrons above certain energies are too dilute to generate
plasma waves. The overall spectrum observed at the Earth becomes close
to a broken power-law, where electrons below the break energy
generate and absorb Langmuir waves, while electrons above the break energy
are not affected by the wave-particle interactions.

\begin{figure} \center\includegraphics[width=120mm]
{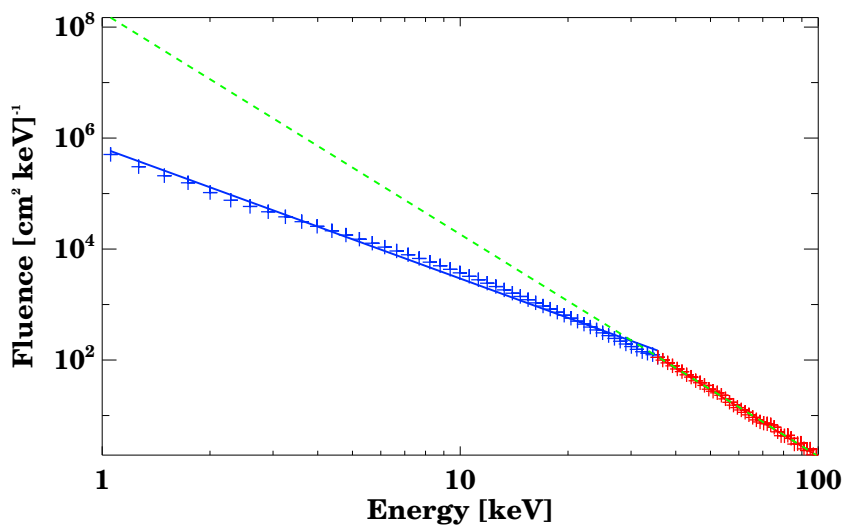}
\caption{The simulated spectrum (fluence [electrons cm$^{-2}$ keV$^{-1}$] )
of solar flare energetic particles at the Earth (crosses). The blue (red) line shows
the power law fit to the spectrum below (above) the break energy $35$ keV.
The green line shows the initially injected fluence.
Spectral index of injected electrons is $\delta_{high}=4$. The spectral index below
the break is $\delta_{low}=2.35$.}
\label{fig:1} \end{figure}

Traditionally {\it in-situ} measurements of energetic electrons
\citep[e.g.][]{Lin_etal95} provide the flux density differential
in energy $F(E,x,t)=f(v,x,t)/m$ [electrons cm$^{-2}$ s$^{-1}$ keV$^{-1}$]
and the fluences (flux integrated over the duration of an event)
[electrons cm$^{-2}$ keV$^{-1}$]. The injected electron fluence in our model
$\int _{-\infty}^{\infty}f(v,x,t)/v dx $ can be calculated from
equations (\ref{init_f},\ref{init_g0}) and is presented in Figure \ref{fig:1}.
The corresponding energy spectral index of the injected electron fluence
at the Sun is $(\alpha +1)/2=\delta+1/2$.  The resulting spectrum of solar energetic
particles at the Earth is also presented in Figure \ref{fig:1}.
As can be seen from Figure \ref{fig:1}, the spectrum of energetic particles
above the break $\sim 35$~keV is identical to the spectrum of injected electrons
so we can deduce these particles have indeed propagated scatter-free
(in our model). The particles below the break energy
do not propagate freely but generate electron plasma waves.
The beam generated plasma waves drift in velocity-space toward lower phase velocities
due to the solar wind density gradient \citep{Kontar01a}. This drift, caused by the decreasing
ambient plasma density, takes waves out of resonance with the particles which generated them
and so reduces the wave energy at a given point in phase space. At lower phase velocities these
waves can be more easily absorbed by the thermal plasma via Landau damping.
Therefore, particles arriving later to this point in phase space are unable
to restore the injected spectrum because they cannot absorb the same amount of energy
from the waves. This results in a flatter energy spectrum of electrons below some energy
where beam-plasma interactions are important (Figure \ref{fig:1}).

\subsection{Break energies and spectral indices}

\begin{figure}
\center\includegraphics[width=80mm]{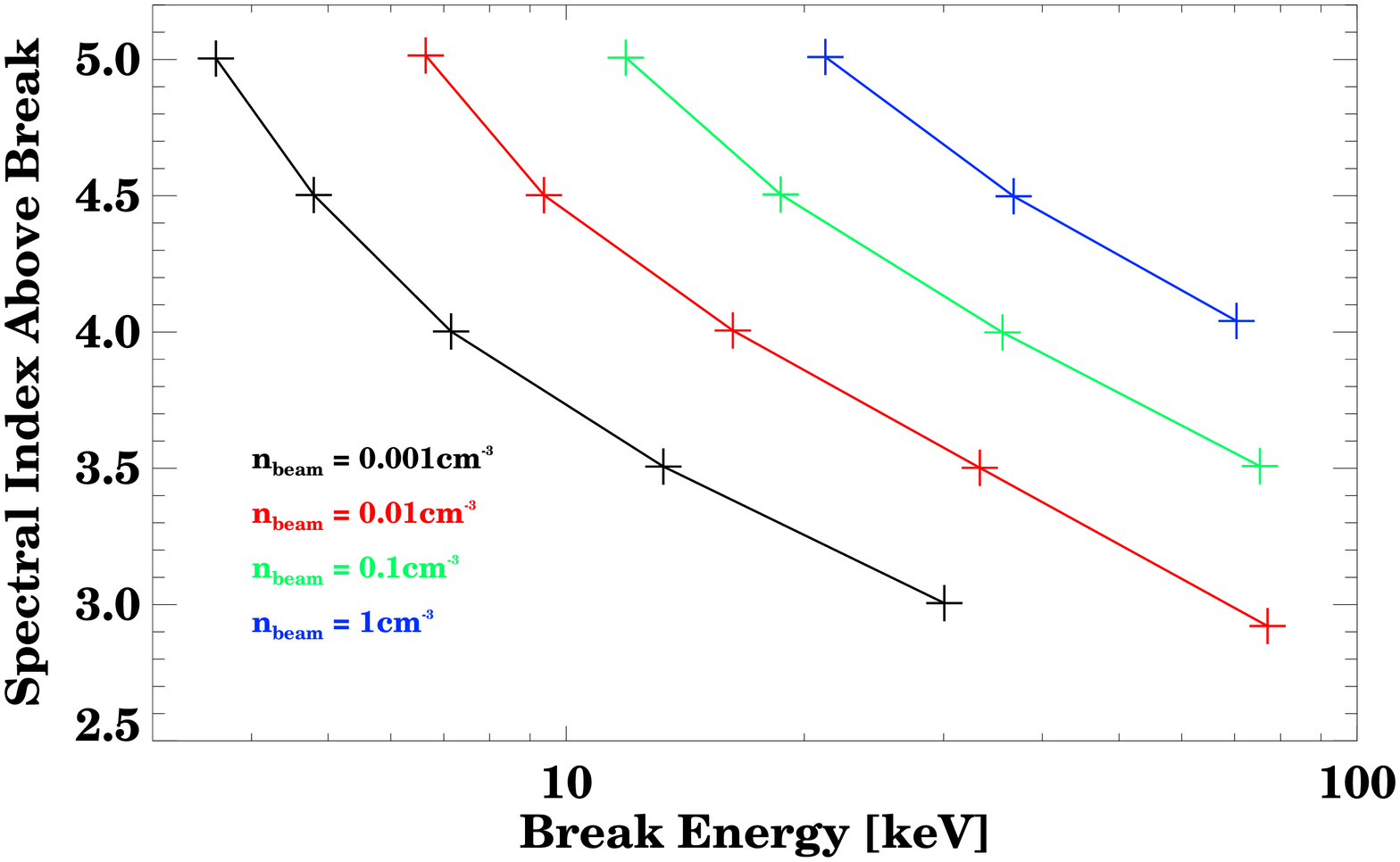}
\center\includegraphics[width=80mm]{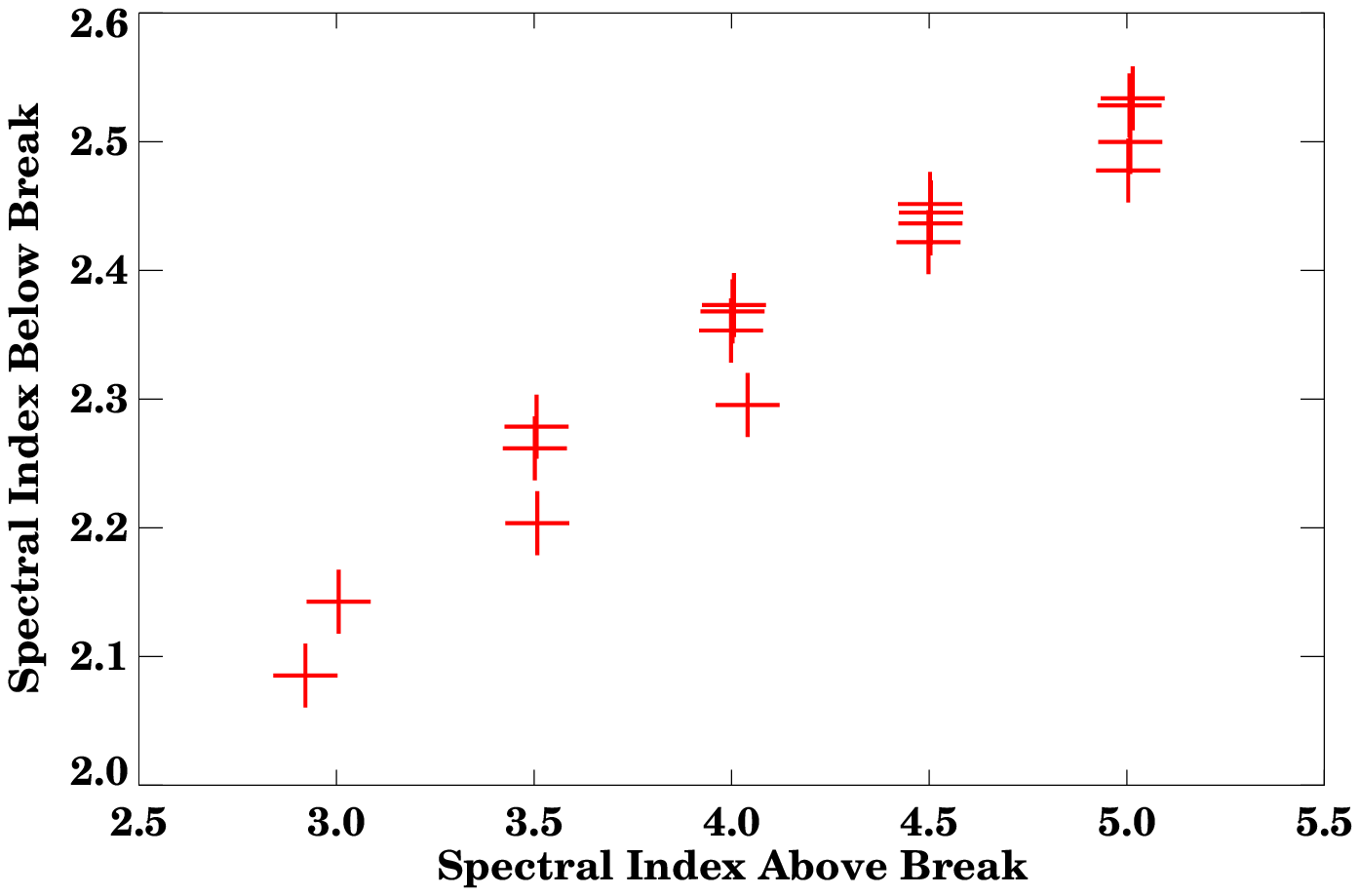}
\center\includegraphics[width=80mm]{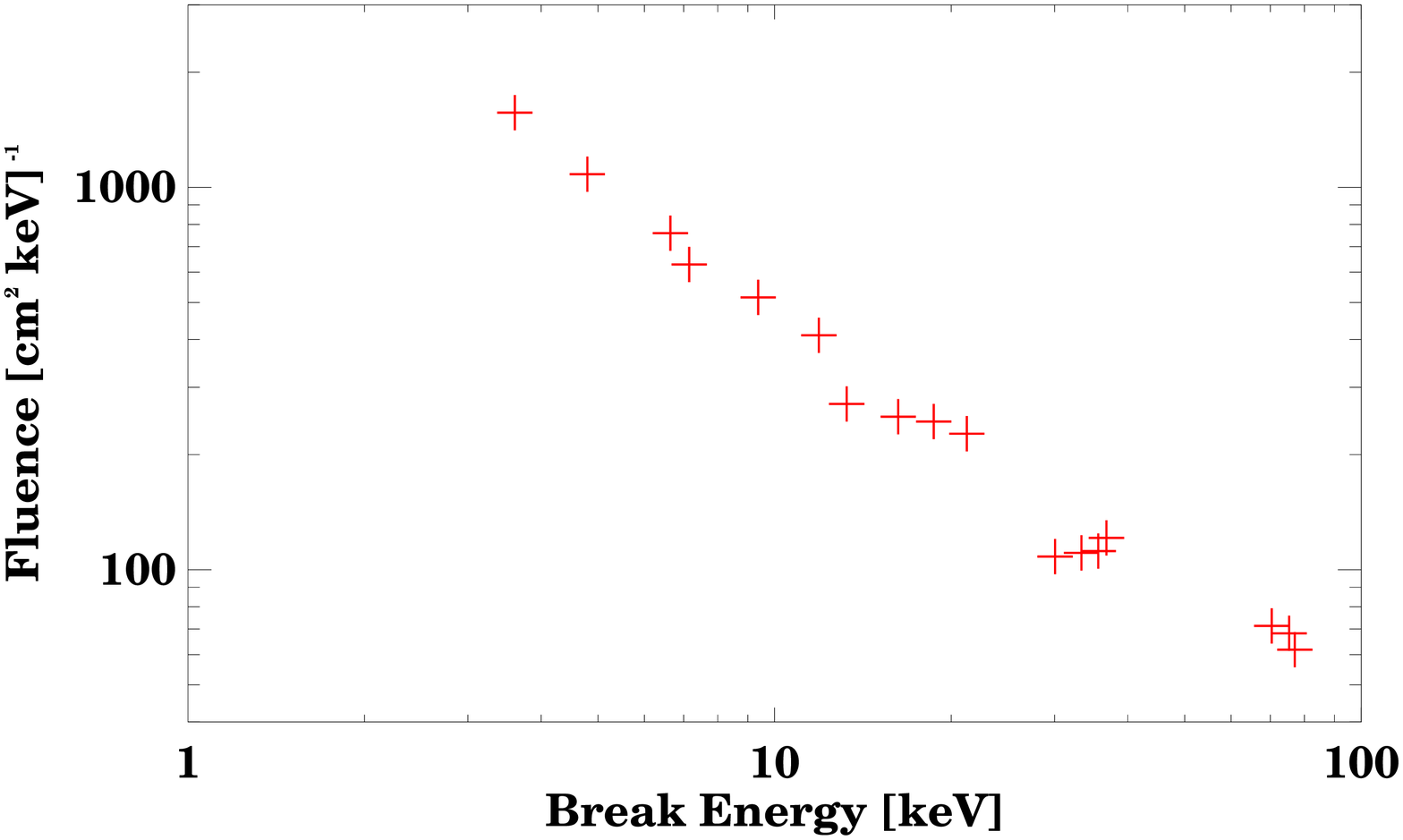}
\caption{Top: Spectral index above the break $\delta_{high}$ versus break energy for various electron
densities and spectral indices;
Middle: Spectral index below the break $\delta_{low}$ energy versus spectral index above the break energy;
Low: Fluence at the break energy versus break energy. Note that the simulations have been done 
for different particle densities and different spectral indices as given in the top panel.}
\label{fig:2}
\end{figure}

We note the spectrum below the break is not exactly a power-law although
resembles one closely, therefore to compare with 
observations, which have normally poor energy resolution
 $\Delta E\sim 0.1 E$ \citep[e.g.][]{Lin_etal95}, we fitted our simulated spectra with simple power-law fits.
The spectral index below the break energy $\delta_{low}$ is
always smaller than the spectral index
above the break energy $\delta_{high}$ and correlates
(Figure \ref{fig:2}) with $\delta_{high}$ in a remarkably similar manner
as observed by \citet{Krucker_etal09}.
The range of $\delta_{low}$ appear in a rather narrow range
between 2 and 2.5 for a wide range of injected spectral
indices between 3 and 5 (Figure \ref{fig:2}).
The actual value of $\delta_{low}$ is also dependent
on the background plasma density and will be different
should the heliospheric density model change.

The break energy range for all simulations is between $4$~keV and $80$~keV
(Figure \ref{fig:2}), with the exact break energy being dependent on the initial spectral
index of the beam, $\delta_{high}$, and the initial density of the beam.
As the initial density of the beam increases, the break of spectral
index occurs at higher energies. Indeed, the larger number of injected electrons,
the faster the generation of plasma waves proceeds and hence the stronger
the interaction between electrons and plasma waves. This also explains the dependence
of break energy to the injected spectral index, with lower spectral indices having
a larger population of higher energy electrons and hence having higher break energies.
The fluence at the break energy correlates to the break energy itself (Figure \ref{fig:2})
with higher break energies corresponding to lower fluence magnitudes.

\subsection{Electron time-of-flight and apparent injection time}

\begin{figure}
\center\includegraphics[width=120mm] {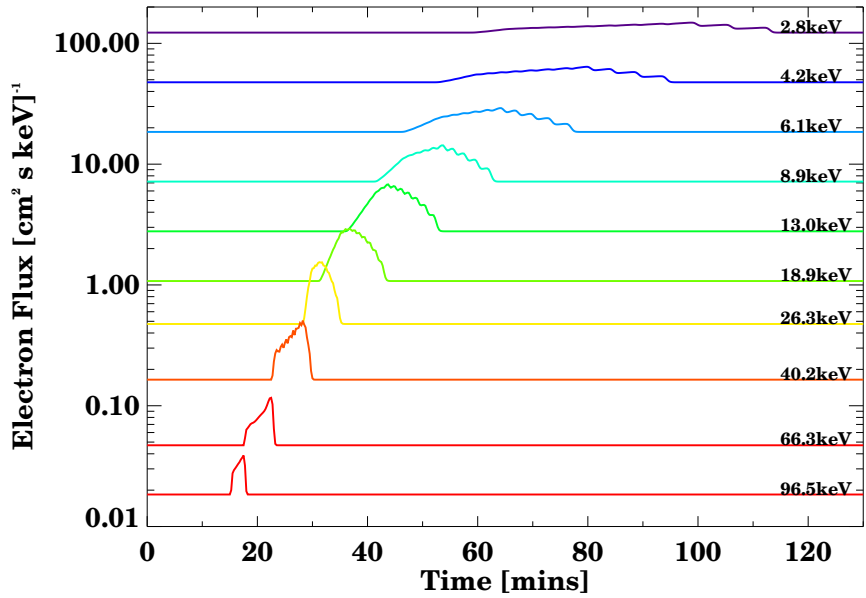}
\caption{Simulated electron flux density time profiles of energetic electrons for WIND/3DP
energies \citep{Lin_etal95}. Electron flux density [electrons cm$^{-2}$ s$^{-1}$ keV$^{-1}$]
as a function of time at $1.2$ AU for 10 energy channels.
The time $t=0$ corresponds to the injection time at the Sun.
The sawtooth structure appearing in low energy channels
is an artifact of finite binning in the velocity space.
The initial beam parameters are the same as in Figure \ref{fig:1}.}
\label{fig:3}
\end{figure}

\begin{figure} \center\includegraphics[width=120mm]
{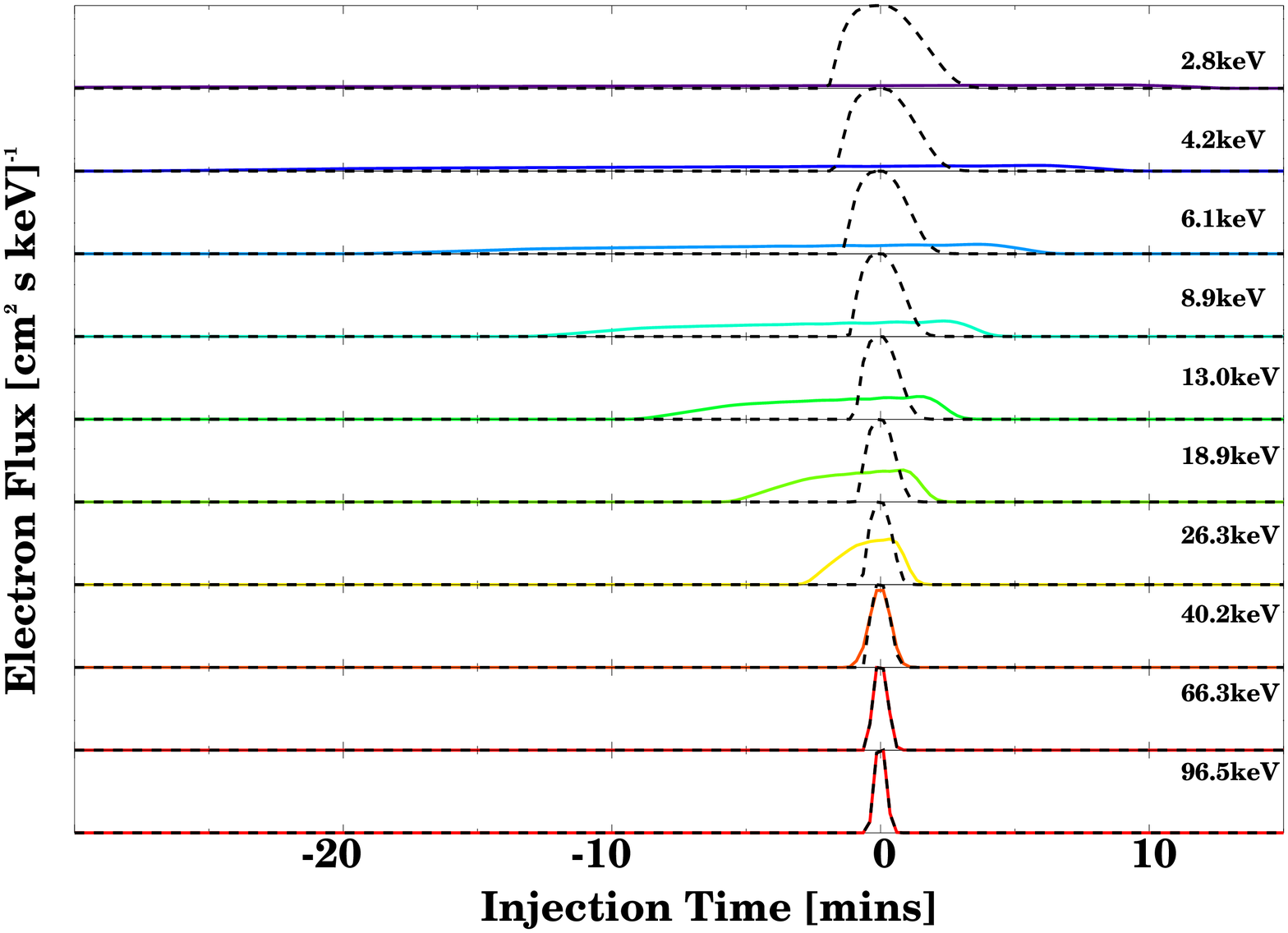}
\caption{The apparent injection profiles of electrons at the Sun {\it assuming}
free streaming of all electrons. The plotted fluxes are normalized to the
maxima of the scatter free case. The injection profiles for scatter-free propagation
(without generation and absorption of waves) is overplotted with dashed lines.
The initial beam parameters are the same as in Figure \ref{fig:3}.
}\label{fig:4}
\end{figure}

The particles arriving at the Earth (1.2 AU) show near
time-of-flight dispersion (Figure \ref{fig:3}), often observed by satellites
in impulsive solar electron events. Assuming scatter-free propagation,
i.e. without any interaction with plasma, one can produce the apparent injection profile
at the Sun, as \citet{Krucker_etal1999,Wang_etal06} did for observations.
Under scatter-free assumption, the time-of-flight for
energetic electrons from the Sun to the spacecraft is simply $t_A=L/v(E)$,
where $L=1.2$~AU is the distance and $v(E)$ is the speed of electrons
for various energies. The time of arrival is then simply $t_A=t_{inj}+L/v(E)$, where
$t_{inj}$ is the apparent injection time. These apparent injection profiles
with background added have been calculated from the simulated fluxes
at 1.2~AU (see Figure \ref{fig:4}). If the electrons propagate scatter-free
they would require $10-20$ minutes earlier onset of injection $t_{inj}$
of low ($3-12$ keV) energy electron injection and a delayed maximum of the
injection to explain observations. An identical simulation was ran with
an electron beam not interacting with the background plasma (scatter free propagation)
and the results are compared in Figure \ref{fig:4}. Comparing the calculated injection
profiles over various
energies one observes good agreement between free-streaming and full simulations
at energies above the break energy 35~keV. The apparent injection profile
at lower energies becomes wider than the free streaming case, and 3-25 keV electrons
show apparent early injection of 1 - 20 minutes before the actual injection.
The lower energy is, the earlier the apparent onset of the injection.
The low energy electrons at the Earth are not only due to the injection at the Sun
but because of the in-flight deceleration of faster particles.
These electrons had initial energy higher than the detected
energy at $1.2$ AU, hence they have traveled a part of the distance
faster than can be inferred from their detected energies. The relaxation of the electron
distribution function towards a flatter shape in velocity space
$\partial f(v,x,t)/\partial v \sim 0$ means at a specific spatial location,
some electrons have energies too low to have arrived by free propagation alone.
In addition, the maximum of apparent injection profile at energies below $\sim 35$ keV appears later.
Therefore, the similar injection profile obtained \citep{Wang_etal06} should be interpreted
as the direct evidence of electron plasma wave scattering in the heliosphere and not the indication
of a separate acceleration mechanism. This early injection time is a direct result
of low-energy electron driven turbulence in non-uniform plasma,
which affects the propagation of electrons.
As evident from Figure \ref{fig:4},
the onset of electron injection is also instrument background dependent
- the higher/lower background levels would lead to later/earlier
injection times for low energy electrons.

\section{Discussion and Conclusions}

The generation and reabsorption of electron plasma waves by the electron
beam in a non-uniform plasma plays an important role in the electron transport and
should be taken into account when in-situ electron measurements are analyzed
and interpreted. The simulations presented successfully reproduce the spectral and
temporal characteristics of observed solar energetic electron events. The scattering
of the solar energetic particles by the beam-driven electrostatic plasma waves
leads to the appearance of a broken power-law in energy spectrum, and the apparent
early injection of low energy electrons in a few keV range at the Sun.
We emphasize here that it is the combined effect of plasma wave generation
and non-uniform density inhomogeneity that leads to the appearance
of a broken power-law.

We have shown that the low energy electrons (below the break)
are originally injected with higher energies but have lost some their energy
to plasma waves in the background plasma
and are therefore detected earlier than their energy at the spacecraft suggest.
The apparent early start of low energy electron injection appears
due to propagation effects and does not require a secondary beam population
\citep[c.f.][]{Wang_etal06}. Moreover, the simulations naturally explain the linear
correlation between low energy and high energy spectral indices often observed
in the spectra of solar impulsive electron events \citep{Krucker_etal09}.

The characteristic time of beam-plasma interaction via electron plasma waves
is inversely proportional to the density of the energetic electrons. If the beam is
dilute, electrons do not generate plasma turbulence and the spectrum of such
electrons could be free from propagation effects (in our model). Such events
are likely to be seen only at low energies \citep[e.g.][]{Krucker_etal09}.
If the beam is dense enough to excite Langmuir waves,
the initially injected power-law spectrum will be detected as a broken power-law.
The break energy is dependent on a number of parameters:
spectral index of injected solar electrons, the density of the energetic electrons,
and the heliospheric density model. Therefore, the correlation
between break energy and fluence at the break energy should be taken with care.
Our simulations also suggest that in order to deduce meaningful time-of-flight
parameters (e.g. injection times), the energetic electrons should be always analyzed
in the channels above the break energy and not above some fixed energy as
sometimes done in the literature.

While our simulations can successfully explain and reproduce a number of
observed properties and highlight the governing role of wave-particle
interactions at keV- tens of keV energies,
the heliospheric plasma - electron beam system is more complex. First,
the different density models \citep[e.g.][]{Newkirk67,Saito_etal1977} could
change the break energy values, although will not alter the overall conclusions.
Second, the heliospheric plasma has density perturbations on various scales
which can affect the evolution of plasma waves \citep{KontarPecseli02},
lead to a spiky structure of the Langmuir waves \citep{Melrose90,Kontar01b,Li_etal2006},
often observed in the interplanetary space \citep{Lin85},
and hence affect the electron distribution.
In addition, the higher energy solar impulsive electron events demonstrate injection
with $\sim 10$~min delay after the onset of type III bursts
\citep[e.g.][]{Krucker_etal1999,HaggertyRoelof02}.
Therefore, additional simulations and in-situ measurements are needed to understand
this complex non-linear system.



\acknowledgments

EPK acknowledges financial support of STFC rolling grant and
STFC/PPARC Advanced Fellowship. Financial support
by the European Commission through the SOLAIRE Network
(MTRN-CT-2006-035484) is gratefully acknowledged.

\bibliographystyle{apj}
\bibliography{refs}

\begin{thebibliography}{35}
\expandafter\ifx\csname natexlab\endcsname\relax\def\natexlab#1{#1}\fi

\bibitem[{{Aschwanden}(2002)}]{Aschwanden02}
{Aschwanden}, M.~J. 2002, Space Science Reviews, 101, 1

\bibitem[{{Benz}(2008)}]{Benz08}
{Benz}, A.~O. 2008, Living Reviews in Solar Physics, 5, 1

\bibitem[{{Brown} \& {Kontar}(2005)}]{BrownKontar05}
{Brown}, J.~C., \& {Kontar}, E.~P. 2005, Advances in Space Research, 35, 1675

\bibitem[{{Cane}(2003)}]{Cane2003}
{Cane}, H.~V. 2003, \apj, 598, 1403

\bibitem[{{Drummond} \& {Pines}(1962)}]{Drummond_Pines1962}
{Drummond}, W.~E., \& {Pines}, D. 1962, Nucl. Fusion Suppl., 3, 1049

\bibitem[{{Dulk}(1985)}]{Dulk85}
{Dulk}, G.~A. 1985, \araa, 23, 169

\bibitem[{{Ergun} {et~al.}(1998){Ergun}, {Larson}, {Lin}, {McFadden},
  {Carlson}, {Anderson}, {Muschietti}, {McCarthy}, {Parks}, {Reme}, {Bosqued},
  {D'Uston}, {Sanderson}, {Wenzel}, {Kaiser}, {Lepping}, {Bale}, {Kellogg}, \&
  {Bougeret}}]{Ergun_etal98}
{Ergun}, R.~E., {Larson}, D., {Lin}, R.~P., {McFadden}, J.~P., {Carlson},
  C.~W., {Anderson}, K.~A., {Muschietti}, L., {McCarthy}, M., {Parks}, G.~K.,
  {Reme}, H., {Bosqued}, J.~M., {D'Uston}, C., {Sanderson}, T.~R., {Wenzel},
  K.~P., {Kaiser}, M., {Lepping}, R.~P., {Bale}, S.~D., {Kellogg}, P., \&
  {Bougeret}, J.-L. 1998, \apj, 503, 435

\bibitem[{{Gaelzer} {et~al.}(2008){Gaelzer}, {Ziebell}, {Vi{\~n}as}, {Yoon}, \&
  {Ryu}}]{2008ApJ...677..676G}
{Gaelzer}, R., {Ziebell}, L.~F., {Vi{\~n}as}, A.~F., {Yoon}, P.~H., \& {Ryu},
  C.-M. 2008, \apj, 677, 676

\bibitem[{{Ginzburg} \& {Zhelezniakov}(1958)}]{GZ1958}
{Ginzburg}, V.~L., \& {Zhelezniakov}, V.~V. 1958, Soviet Astronomy, 2, 653

\bibitem[{{Gosling} {et~al.}(2003){Gosling}, {Skoug}, \&
  {McComas}}]{Gosling_etal2003}
{Gosling}, J.~T., {Skoug}, R.~M., \& {McComas}, D.~J. 2003, \grl, 30, 130000

\bibitem[{{Grognard}(1982)}]{Grognard1982}
{Grognard}, R.~J.-M. 1982, \solphys, 81, 173

\bibitem[{{Haggerty} \& {Roelof}(2002)}]{HaggertyRoelof02}
{Haggerty}, D.~K., \& {Roelof}, E.~C. 2002, \apj, 579, 841

\bibitem[{{Kontar}(2001{\natexlab{a}})}]{Kontar01a}
{Kontar}, E.~P. 2001{\natexlab{a}}, \solphys, 202, 131

\bibitem[{{Kontar}(2001{\natexlab{b}})}]{Kontar01b}
---. 2001{\natexlab{b}}, \aap, 375, 629

\bibitem[{{Kontar} \& {P{\'e}cseli}(2002)}]{KontarPecseli02}
{Kontar}, E.~P., \& {P{\'e}cseli}, H.~L. 2002, \pre, 65, 066408

\bibitem[{{Krucker} {et~al.}(2007){Krucker}, {Kontar}, {Christe}, \&
  {Lin}}]{Krucker_etal07}
{Krucker}, S., {Kontar}, E.~P., {Christe}, S., \& {Lin}, R.~P. 2007, \apjl,
  663, L109

\bibitem[{{Krucker} {et~al.}(1999){Krucker}, {Larson}, {Lin}, \&
  {Thompson}}]{Krucker_etal1999}
{Krucker}, S., {Larson}, D.~E., {Lin}, R.~P., \& {Thompson}, B.~J. 1999, \apj,
  519, 864

\bibitem[{{Krucker} {et~al.}(2009){Krucker}, {Oakley}, \&
  {Lin}}]{Krucker_etal09}
{Krucker}, S., {Oakley}, P.~H., \& {Lin}, R.~P. 2009, \apj, 691, 806

\bibitem[{{Ledenev} {et~al.}(2004){Ledenev}, {Zverev}, \&
  {Starygin}}]{Ledenev04}
{Ledenev}, V.~G., {Zverev}, E.~A., \& {Starygin}, A.~P. 2004, \solphys, 222,
  299

\bibitem[{{Li} {et~al.}(2006){Li}, {Robinson}, \& {Cairns}}]{Li_etal2006}
{Li}, B., {Robinson}, P.~A., \& {Cairns}, I.~H. 2006, Physical Review Letters,
  96, 145005

\bibitem[{{Lin}(1985)}]{Lin85}
{Lin}, R.~P. 1985, \solphys, 100, 537

\bibitem[{{Lin} {et~al.}(1995){Lin}, {Anderson}, {Ashford}, {Carlson},
  {Curtis}, {Ergun}, {Larson}, {McFadden}, {McCarthy}, {Parks}, {R{\`e}me},
  {Bosqued}, {Coutelier}, {Cotin}, {D'Uston}, {Wenzel}, {Sanderson}, {Henrion},
  {Ronnet}, \& {Paschmann}}]{Lin_etal95}
{Lin}, R.~P., {Anderson}, K.~A., {Ashford}, S., {Carlson}, C., {Curtis}, D.,
  {Ergun}, R., {Larson}, D., {McFadden}, J., {McCarthy}, M., {Parks}, G.~K.,
  {R{\`e}me}, H., {Bosqued}, J.~M., {Coutelier}, J., {Cotin}, F., {D'Uston},
  C., {Wenzel}, K.-P., {Sanderson}, T.~R., {Henrion}, J., {Ronnet}, J.~C., \&
  {Paschmann}, G. 1995, Space Science Reviews, 71, 125

\bibitem[{{Lin} {et~al.}(1996){Lin}, {Larson}, {McFadden}, {Carlson}, {Ergun},
  {Anderson}, {Ashford}, {McCarthy}, {Parks}, {R{\`e}me}, {Bosqued}, {d'Uston},
  {Sanderson}, \& {Wenzel}}]{Lin_etal1996}
{Lin}, R.~P., {Larson}, D., {McFadden}, J., {Carlson}, C.~W., {Ergun}, R.~E.,
  {Anderson}, K.~A., {Ashford}, S., {McCarthy}, M., {Parks}, G.~K., {R{\`e}me},
  H., {Bosqued}, J.~M., {d'Uston}, C., {Sanderson}, T.~R., \& {Wenzel}, K.~P.
  1996, \grl, 23, 1211

\bibitem[{{Lin} {et~al.}(1981){Lin}, {Potter}, {Gurnett}, \&
  {Scarf}}]{Li_etal1981}
{Lin}, R.~P., {Potter}, D.~W., {Gurnett}, D.~A., \& {Scarf}, F.~L. 1981, \apj,
  251, 364

\bibitem[{{Mann} {et~al.}(1999){Mann}, {Jansen}, {MacDowall}, {Kaiser}, \&
  {Stone}}]{Mann_etal1999}
{Mann}, G., {Jansen}, F., {MacDowall}, R.~J., {Kaiser}, M.~L., \& {Stone},
  R.~G. 1999, \aap, 348, 614

\bibitem[{{Melnik} {et~al.}(1999){Melnik}, {Lapshin}, \& {Kontar}}]{Melnik99}
{Melnik}, V.~N., {Lapshin}, V., \& {Kontar}, E. 1999, \solphys, 184, 353

\bibitem[{{Melrose}(1985)}]{Melrose_85}
{Melrose}, D.~B. 1985, {Plasma emission mechanisms} (Solar Radiophysics:
  Studies of Emission from the Sun at Metre Wavelengths), 177--210

\bibitem[{{Melrose}(1990)}]{Melrose90}
---. 1990, \solphys, 130, 3

\bibitem[{{Newkirk}(1967)}]{Newkirk67}
{Newkirk}, G.~J. 1967, \araa, 5, 213

\bibitem[{{Parker}(1958)}]{parker58}
{Parker}, E.~N. 1958, \apj, 128, 664

\bibitem[{{Saito} {et~al.}(1977){Saito}, {Poland}, \& {Munro}}]{Saito_etal1977}
{Saito}, K., {Poland}, A.~I., \& {Munro}, R.~H. 1977, \solphys, 55, 121

\bibitem[{{van Allen} \& {Krimigis}(1965)}]{vanAllen65}
{van Allen}, J.~A., \& {Krimigis}, S.~M. 1965, \jgr, 70, 5737

\bibitem[{{Vedenov} {et~al.}(1962){Vedenov}, {Lelikhov}, \&
  {Sagdeev}}]{Vedenov_etal1962}
{Vedenov}, A.~A., {Lelikhov}, E.~P., \& {Sagdeev}, R.~Z. 1962, Nucl. Fusion
  Suppl., 2, 465

\bibitem[{{Wang} {et~al.}(2006){Wang}, {Lin}, {Krucker}, \&
  {Gosling}}]{Wang_etal06}
{Wang}, L., {Lin}, R.~P., {Krucker}, S., \& {Gosling}, J.~T. 2006, \grl, 33,
  3106

\bibitem[{{Zaitsev} {et~al.}(1972){Zaitsev}, {Mityakov}, \&
  {Rapoport}}]{Zaitsev_etal72}
{Zaitsev}, V.~V., {Mityakov}, N.~A., \& {Rapoport}, V.~O. 1972, \solphys, 24,
  444

\end{thebibliography}

\clearpage


\end{document}